\documentclass[twocolumn,pre,twoside,showpacs,showkeys,preprintnumbers,floatfix]{revtex4}

\usepackage{amsmath}
\usepackage{amssymb}
\usepackage{graphicx}
\usepackage{dcolumn}
\usepackage{bm}

\def\m#1{\mathrm{#1}}
\def\Ref#1{(\ref{eq:#1})}

\def\p{\partial}
\def\epsilon{\varepsilon}
\def\theta{\vartheta}
\def\rho{\varrho}

\def\Int#1#2{\int\!\mathrm{d}^{#1}{#2}\;}
\def\set#1{\underline{#1}}

\begin{document}


\title{Nonequilibrium steady states in fluids of platelike colloidal particles}

\author{Markus Bier}
\email{m.bier@phys.uu.nl}

\author{Ren\'e van Roij}

\affiliation
{
   Institute for Theoretical Physics, 
   Utrecht University, 
   Leuvenlaan 4, 
   3584CE Utrecht, 
   The Netherlands
}

\date{October 29, 2007}

\begin{abstract}
Nonequilibrium steady states in an open system connecting two reservoirs of platelike colloidal particles are 
investigated by means of a recently proposed phenomenological dynamic density functional theory [M.\ Bier and 
R.\ van Roij, Phys.\ Rev.\ E \textbf{76}, 021405 (2007)]. The platelike colloidal particles are approximated 
within the Zwanzig model of restricted orientations, which exhibits an isotropic-nematic bulk phase transition. 
Inhomogeneities of the local chemical potential generate a diffusion current which relaxes to a nonvanishing 
value if the two reservoirs coupled to the system sustain different chemical potentials. The relaxation process 
of initial states towards the steady state turns out to comprise two regimes: a smoothening of initial 
steplike structures followed by an ultimate relaxation of the slowest diffusive mode. The position of a 
nonequilibrium interface and the particle current of steady states depend nontrivially on the structure of the 
reservoirs due to the coupling between translational and orientational degrees of freedom of the fluid.
\end{abstract}


\maketitle


\section{Introduction}
\label{sec:introduction}

Complex fluids of platelike colloidal particles, e.g., clay suspensions, have been investigated to quite an extent
in recent years. The scientific interest largely stems from the enormous range of phenomena found in these systems
such as flocculation, glass transitions, gelation, aging, and even liquid crystal phase transitions 
\cite{Mourchid1995,Brown1998,Mourchid1998,vanderKooij1998,Bonn1999,Brown1999,Levitz2000,Knaebel2000,Abou2001,%
vanderBeek2003,Liu2003,vanderBeek2004,Wang2005} due to the interplay of translational \emph{and} orientational 
degrees of freedom of the 
constituting particles. Whereas the liquid crystal properties of fluids of platelike colloidal particles have been 
addressed in some theoretical studies devoted to homogeneous as well as inhomogeneous \emph{equilibrium} systems
\cite{Cuesta1999,Rowan2002,Harnau2001,Harnau2002a,Harnau2002b,Bier2004,Harnau2004,Costa2005,Harnau2005,%
Bier2005,Bier2006,vanderBeek2006,Reich2007}, not many investigations of the \emph{nonequilibrium} behavior have 
been performed.

In a recent publication the authors proposed a phenomenological \emph{dynamic density functional theory} (DDFT)
in order to describe the relaxation dynamics of fluids of platelike colloidal particles under the influence of
an external field \cite{Bier2007}. Relaxation into stable or metastable states, both characterized by a homogeneous
local chemical potential, has been found depending on the initial state and the external field. In the present
contribution, applying the same DDFT formalism, \emph{nonequilibrium steady states} are investigated which form 
within a channel connecting two particle reservoirs of \emph{different} chemical potentials. As 
the local chemical potential in the channel is expected to be spatially \emph{inhomogeneous} DDFT can be considered 
as the natural formalism to describe the fluid of platelike colloidal particles on a coarsegrained level.

DDFT is an extension of equilibrium \emph{density functional theory} (DFT) \cite{Evans1979,Evans1989,Evans1991} to 
nonequilibrium conditions by proposing an equation of motion for the one-particle densities \cite{Dieterich1990}. 
The one-particle densities are \emph{assumed} to describe the (dynamical) state of the system completely. 
On the one hand, the DDFT equations are similar to the traditional time-dependent Landau-Ginzburg and Cahn-Hillard 
models of critical dynamics, spinodal decomposition, and crystal growth \cite{Langer1971,Kawasaki1977,Collins1985,%
Harrowell1987,Boettinger2002,Granasy2006}. On the other hand, the DDFT equations have been derived 
within the framework of (overdamped) Langevin dynamics, which is considered a reasonable description for dilute
colloidal dispersions \cite{Dean1996,Marconi1999,Marconi2000,Archer2004}. As the number densities of the present 
work are close to the isotropic-nematic two-phase region, which is located at small densities for highly 
anisotropic particles, the applied DDFT, which neglects hydrodynamic interactions, is expected to be valid. The 
DDFT proposed in Ref.~\cite{Bier2007} is briefly summarized in Sec.~\ref{sec:formalism}.

The present work is restricted to \emph{fluids} of platelike colloidal particles, i.e., nonergodic states 
such as glasses or gels are beyond the scope of this contribution. It is known that an ergodic system described 
by a linear master equation relaxes towards a unique steady state \cite{Schnakenberg1976}. Although the formalism
applied here is not equivalent to a linear master equation, the relaxation towards a unique steady state
is nonetheless expected. This relaxation process is described in Sec.~\ref{sec:relaxation} in terms of the number
density, the orientational order, the local chemical potential and the particle current.

The final steady state is necessarily a \emph{nonequilibrium} steady state, because the particle reservoirs 
coupled to the system preclude equilibration of the system. This nonequilibrium steady state exhibits a 
nonvanishing particle current which is sustained by the chemical potential difference of the particle reservoirs
coupled to the system. Moreover, for suitably chosen reservoir chemical potentials, the nonequilibrium steady state
shows signs of bulk phase transitions. These issues are discussed in Sec.~\ref{sec:properties}.

Section~\ref{sec:discussion} discusses the results found in Secs.~\ref{sec:relaxation} and \ref{sec:properties} and
closes with a short summary.


\section{Formalism}
\label{sec:formalism}


\subsection{Model fluid and system geometry}
\label{subsec:model}

Consider a dispersion of monodisperse, hard, infinitely thin, square colloidal particles within a three-dimensional
continuous solvent. The side length of the square particles is $D$. The orientations, described by the normal 
vector of the square face, are restricted to directions parallel to the Cartesian axes (Zwanzig approximation 
\cite{Zwanzig1963}). A particle is called an $i$ particle if its orientation is along the $i$ axis, $i\in\{x,y,z\}$. 

The system under consideration is a channel of length $H$ which connects two particle reservoirs. The channel is 
assumed to be much wider than the particle size $D$ such that effects of the channel walls onto the colloidal 
fluid are negligible. Consequently the fluid structure in the channel is expected to vary only along the channel 
axis which is taken as the $z$ axis. The channel is located at the $z$ axis interval $[0,H]$. The local number 
density of $i$ particles in the channel at position $z\in[0,H]$ is denoted by $\rho_i(z)$. The abbreviation 
$\set{\rho}:=(\rho_x,\rho_y,\rho_z)$ is used later. 

The structure of the model fluid is adequately described in terms of the \emph{total density} $\rho:=\sum_i\rho_i$
and the \emph{order parameter tensor} $Q$ \cite{deGennes1993} which, within Zwanzig
models, is given by
\begin{equation}
   Q_{ii'} = \frac{1}{2}\left(3\frac{\rho_i}{\rho}-1\right)\delta_{ii'},
   \label{eq:Q}
\end{equation}
where $\delta_{ii'}$ is the Kronecker delta.

The two particle reservoirs are connected to the channel at positions $z=0$ and $z=H$, respectively. They are assumed
to sustain equilibrium bulk structures $\set{\rho}_0$ and $\set{\rho}_H$ corresponding to the chemical potentials 
$\mu_0$ and $\mu_H$, respectively. The coupling of the two reservoirs to the system amounts to the 
Dirichlet-like boundary conditions $\set{\rho}(z \leq 0,t)=\set{\rho}_0$ and $\set{\rho}(z \geq H,t)=\set{\rho}_H$.


\subsection{Dynamic density functional theory}
\label{subsec:ddft}

Within this work the dynamic density functional theory (DDFT) proposed by the authors in Ref.~\cite{Bier2007}
is applied. It consists of the set of equations of motion for the one-particle densities $\rho_i$ given by
\begin{equation}
   \frac{\p \rho_i(z,t)}{\p t} =
   \bigg(\frac{\p \rho_i(z,t)}{\p t}\bigg)_\m{trans} +
   \bigg(\frac{\p \rho_i(z,t)}{\p t}\bigg)_\m{rot},
   \label{eq:eom}
\end{equation}
with the translational diffusion for fixed orientation described by
\begin{equation}
   \bigg(\frac{\p \rho_i(z,t)}{\p t}\bigg)_\m{trans} :=
   -\frac{\partial j_i(z,[\set{\rho}(t)])}{\partial z},
   \label{eq:transpart}
\end{equation}
where the particle currents along the $z$ axis are
\begin{equation}
   j_i(z,[\set{\rho}]) :=
   -\Gamma_i\rho_i(z)\frac{\partial\beta\mu_i(z,[\set{\rho}])}{\partial z},
   \label{eq:diffcurr}
\end{equation}
and with the rotational diffusion at fixed position $z$ described by
\begin{eqnarray}
   \bigg(\frac{\p \rho_i(z,t)}{\p t}\bigg)_\m{rot} 
   & := & 
   -\frac{1}{6\tau}\sum_{i'}\big(\rho_i(z,t)+\rho_{i'}(z,t)\big)
   \label{eq:rotpart}\\
   & &
   \Big(\beta\mu_i\big(z,[\set{\rho}(t)]\big)-\beta\mu_{i'}\big(z,[\set{\rho}(t)]\big)\Big).
   \nonumber
\end{eqnarray}
The driving force is due to inhomogeneities of the \emph{local chemical potential}
\begin{equation}
   \mu_i\big(z,[\set{\rho}]\big) := \frac{\delta F}{\delta\rho_i(z)}\bigg|_{\set{\rho}},
   \label{eq:locchempot}
\end{equation}
which derives from the free energy functional \cite{Cuesta1997.1,Cuesta1997.2}
\begin{eqnarray}
   \beta F[\set{\rho}] 
   &\!\!=\!\!&
   \Int{}{z}\bigg(
   \sum_i\rho_i(z)\left(\ln(\rho_i(z)\Lambda^3)-1\right) + 
   \nonumber\\
   & &
   \hphantom{\Int{}{z}\bigg(}
   \Phi(\set{n}(z))\bigg),
   \label{eq:freeenergy}
\end{eqnarray}
where 
\begin{eqnarray}
   \Phi(\set{n}(z)) & = & 
   n_0(z)\ln(1-n_3(z)) + 
   \nonumber\\
   & &
   \frac{\displaystyle\sum_q n_{1q}(z)n_{2q}(z)}{1-n_3(z)} + 
   \frac{\displaystyle\prod_q n_{2q}(z)}{(1-n_3(z))^2}
\end{eqnarray}
describes the excess free energy density due to the hard-core interaction. The weighted densities 
$n_\alpha(z) := \sum_i \omega_{\alpha,i}\otimes\rho_i(z)$ involve the convolution ($\otimes$) of the
densities $\rho_i$ with the weight functions
\begin{eqnarray}
   \omega_{0,i}(z)  & = & a(z,S_{zi}),
   \nonumber\\
   \omega_{1x,i}(z) & = & S_{xi}a(z,S_{zi}),  
   \nonumber\\
   \omega_{1y,i}(z) & = & S_{yi}a(z,S_{zi}),   
   \nonumber\\
   \omega_{1z,i}(z) & = & b(z,S_{zi}),   
   \nonumber\\
   \omega_{2x,i}(z) & = & S_{yi}b(z,S_{zi}),   
   \nonumber\\
   \omega_{2y,i}(z) & = & S_{xi}b(z,S_{zi}),   
   \nonumber\\
   \omega_{2z,i}(z) & = & S_{xi}S_{yi}a(z,S_{zi}),   
   \nonumber\\
   \omega_{3,i}(z)  & = & S_{xi}S_{yi}b(z,S_{zi}),   
   \label{eq:weightfunctions}
\end{eqnarray}
where the abbreviations $a(z,S):=\frac{1}{2}(\delta(\frac{S}{2}+z)+\delta(\frac{S}{2}-z))$ and
$b(z,S):=\Theta(\frac{S}{2}-|z|)$ are used and $S_{qi}\in\{0,D\}$ denotes the extension of $i$ particles
along the $q$ axis. The \emph{rotational relaxation time}
\begin{equation}
   \tau = \frac{2}{9} \beta\eta D^3,
   \label{eq:tau}
\end{equation}
and the \emph{translational diffusion constants}
\begin{equation}
   \Gamma_{x,y} = \frac{D^2}{24\tau}, \quad \Gamma_z = \frac{D^2}{36\tau}
   \label{eq:Gamma}
\end{equation}
have been chosen \cite{Brenner1974}. In the above equations $\beta$ is the inverse temperature, 
$\Lambda$ denotes the thermal de Broglie wavelength, and $\eta$ is the viscosity of the solvent.

The described DDFT neglects hydrodynamic interactions between the platelike particles. This is considered
as a reasonable approximation because of the small particle densities in this study close to the bulk 
isotropic-nematic two-phase coexistence region \cite{Qiu1990,Xue1992}.

An equilibrium state $\set{\rho}^\m{eq}$ fulfills the \emph{Euler-Lagrange equation}
\begin{equation}
   \mu_i\big(z,[\set{\rho}^\m{eq}]\big) = \mu,
   \label{eq:ele}
\end{equation}
with the chemical potential $\mu$, i.e., \emph{equilibrium} density profiles render the local chemical 
potential as a function of position ($z$) and orientation ($i$) into a constant.
Therefore the described DDFT is consistent with equilibrium DFT because any equilibrium state $\set{\rho}^\m{eq}$ 
is stationary under the dynamics represented by Eqs.~\Ref{eom}--\Ref{rotpart}. Moreover, from Eqs.~\Ref{diffcurr} 
and \Ref{ele} one concludes a vanishing particle current for equilibrium states: $j[\set{\rho}^\m{eq}]=0$.

For later convenience, the \emph{reduced (local) chemical potential} 
$\mu^* := \beta\mu - 3\log(\frac{2\Lambda}{D})$ is defined. 


\subsection{Bulk phase behavior and numerical method}
\label{subsec:bulk}

The described model of fluids of platelike colloidal particles exhibits a first-order isotropic-nematic bulk phase 
transition at a reduced chemical potential $\mu_\m{b}^* = -1.3727$. At this binodal (b) the iso\-tro\-pic bulk 
phase of density $\rho_\m{b}^\m{iso}D^3 = 1.1440$ coexists with the nematic bulk phase of density 
$\rho_\m{b}^\m{nem}D^3 = 1.5789$ and scalar order parameter 
$S^\m{nem}_\m{b} = \langle\frac{3}{2}\cos(\theta)^2-\frac{1}{2}\rangle = 0.82696$, where $\langle\cdot\rangle$ 
denotes the thermal average and $\theta$ is the angle between the particle orientation and the director 
\cite{deGennes1993}.

Solutions of the DDFT equation specified in Sec.~\ref{subsec:ddft} are calculated numerically by means of the 
Euler-forward method with integration time steps of $\frac{\tau}{20}$.


\section{Relaxation towards the steady state}
\label{sec:relaxation}

In this section the relaxation process of an initial state $\set{\rho}(t=0)$ towards the stationary state 
corresponding to the boundary conditions exerted by the two reservoirs coupled to the system at $z=0$ and $z=H$
(see Sec.~\ref{sec:formalism}) is studied. Here the arbitrary case of the reservoir at $z=0$ sustaining the
\emph{isotropic} bulk structure $\set{\rho}_0$ corresponding to the reduced chemical potential $\mu^*_0=-1.9$ and 
the reservoir at $z=H$ sustaining the \emph{nematic} bulk structure with the director \emph{parallel} to the $z$ 
axis $\set{\rho}_H$ corresponding to $\mu^*_H=-1.1$ is described in detail. Moreover, the initial state discussed 
here is chosen as
\begin{equation}
   \set{\rho}(z,t=0) =
   \left\{\begin{array}{ll}
      \set{\rho}_H & , z > \frac{H}{2} \\
      \set{\rho}_0 & , z < \frac{H}{2}
   \end{array}\right.,
   \label{eq:init}
\end{equation}
which approximates the equilibrium structure in the channel in the presence of an impermeable membrane located
at $z=\frac{H}{2}$. Here a detailed discussion of the case of the nematic state $\set{\rho}_H$ with the director 
\emph{perpendicular} to the $z$ axis is not necessary, because the temporal evolution turns out to be 
qualitatively the same as for the case of \emph{parallel} alignment.

The solutions of the DDFT equations Eq.~\Ref{eom}--\Ref{Gamma} have also been calculated for different initial 
states than Eq.~\Ref{init}, such as the linear density profile 
$\set{\rho}(0 \leq z \leq H,t=0)=(1-\frac{z}{H})\set{\rho}_0+\frac{z}{H}\set{\rho}_H$ and the almost empty channel 
$\set{\rho}(0 \leq z \leq H,t=0) \approx 0$, in order to verify that the final steady state is independent of the 
initial state.

\begin{figure*}[!t]
   \includegraphics[width=17cm]{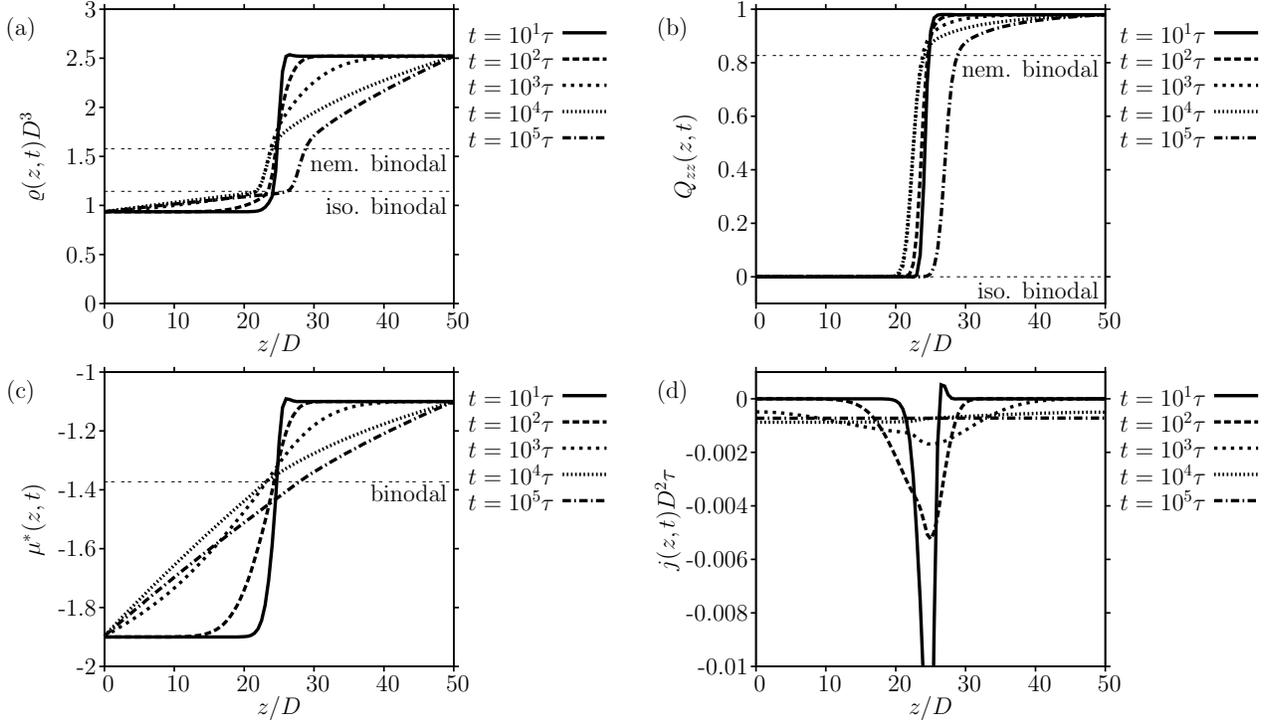}
   \caption{\label{fig:A} (a) Total number density $\rho(z,t)$, (b) nematic order paramter tensor component
           parallel to the $z$ axis $Q_{zz}(z,t)$, (c) reduced local chemical potential $\mu^*(z,t)$, and 
           (d) total particle current $j(z,t)$ of platelike colloidal particles of size $D$ in a 
           channel of length $H=50D$ connecting two particle reservoirs which sustain bulk equilibrium structures 
           corresponding to the reduced chemical potentials $\mu^*_0=-1.9$ at $z=0$ and $\mu^*_H=-1.1$ at $z=H$. 
           The structure in the reservoir at $z=0$ is isotropic whereas the structure in the reservoir at $z=H$ is 
           nematic with the director parallel to the $z$ axis. The isotropic and the nematic binodals of the 
           colloidal fluid in equilibrium are displayed for comparison. The time is given in terms of the 
           rotational relaxation time $\tau$. No further changes are visible for times $t \geq 10^5\tau$.}
\end{figure*}

In Fig.~\ref{fig:A} the temporal evolution of the initial state Eq.~\Ref{init} for the channel
length $H=50D$ is displayed.

The total number density $\rho(z,t)$ is shown in Fig~\ref{fig:A}a. Up to time $t \lessapprox 10^4\tau$ the initial
steplike density profile smoothens by diffusion until the whole channel at $z\in[0,H]$ is affected. $\rho(z,t)$
for times $t \geq 10^5\tau$ does not undergo visible changes, i.e., from then on the steady state is (practically) 
attained. The equilibrium bulk density values of the isotropic binodal $\rho^\m{iso}_\m{b}$ and the nematic
binodal $\rho^\m{nem}_\m{b}$ are shown in Fig.~\ref{fig:A}a, too. The steady state is approximately linear below
the isotropic and above the nematic binodal densities. In the density range between the binodals, corresponding 
to the bulk two-phase coexistence region, a steep portion of the steady state density profile is related to the
isotropic-nematic bulk phase transition. It has been verified by solving the DDFT equations that in the case of
two reservoirs of both isotropic structure a similar steep portion is absent. 

Figure~\ref{fig:A}b displays the temporal evolution of the nematic order parameter tensor component parallel to
the $z$ axis $Q_{zz}(z,t)$. Similar to the total number density profile $\rho(z,t)$ in Fig.~\ref{fig:A}a, a
diffusive smoothening of the initially steplike profile takes place at times $t \lessapprox 10^4\tau$ whereas the
steady state has been attained at times $t \geq 10^5\tau$. Between the order parameter values at the isotropic binodal
$0$ and the nematic binodal $S^\m{nem}_\m{b}$ the order parameter profile increases rapidly as a function of
position.

For times larger than the rotational relaxation time, $t \gg \tau$, the local chemical potentials $\mu^*_i$ for
$i$ particles are practically independent of $i$. Hence it is useful to consider the reduced chemical potential 
profile $\mu^*(z,t) := \frac{1}{3}\sum_i\mu^*_i(z,t)$, which is depicted in 
Fig.~\ref{fig:A}c. Again a smoothening of the initial steplike profile in the time range $t \lessapprox 10^4\tau$
due to diffusion is followed by a restructuring into the final steady state which is has been reached at times
$t \geq 10^5\tau$. The slope of the steady state profile as a function of position $z$ slightly decreases upon
increasing $z$.

In Fig.~\ref{fig:A}d the total particle current $j(z,t) := \sum_i j_i(z,t)$ is displayed. At early times the 
current is localized near the position $z=\frac{H}{2}$ of the discontinuity of the initial state
Eq.~\Ref{init}. Because $\mu^*_0 < \mu^*_H$ the current is \emph{negative}. With time the spatial
current distribution $j(z,t)$ broadens and ultimately becomes homogeneous in the steady state.
Under the present conditions (reservoir chemical potentials, nematic director alignment, channel length) the
steady state current $j(z,t=\infty)D^2\tau=-7.219\cdot 10^{-4}$ is attained.

In order to quantify the ''distance'' of a given state $\set{\rho}$ from the steady state, which is characterized
by a homogeneous particle current, consider the \emph{nonstationarity parameter}
\begin{equation}
   \epsilon[\set{\rho}] := \max_{z\in[0,H]}j(z,[\set{\rho}]) - \min_{z\in[0,H]}j(z,[\set{\rho}]),
   \label{eq:epsilon}
\end{equation}
which is a functional of the state $\set{\rho}$. Obviously $\epsilon[\set{\rho}] \geq 0$. The nonstationarity
parameter vanishes if and only if the current is homogeneous, e.g., if $\set{\rho}$ is a steady state.

\begin{figure}[!t]
   \includegraphics[width=8.5cm]{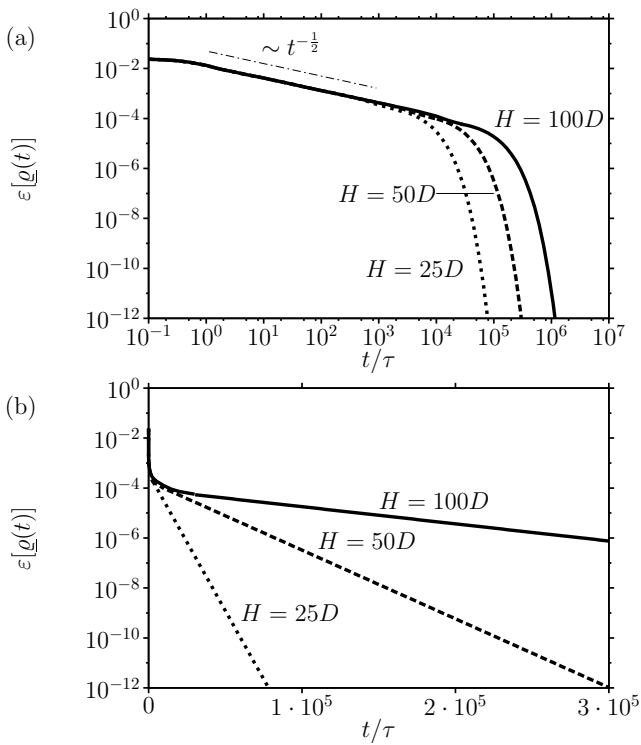}
   \caption{\label{fig:B}Nonstationarity parameter $\epsilon[\set{\rho}(t)]$, defined in the main text, of state 
           $\set{\rho}(t)$ at time $t$ for channel lengths $H=100D$ (solid line), $H=50D$ (dashed line), and $H=25D$
           (dotted line). The time is given in terms of the rotational relaxation time $\tau$. In the time range 
           $\tau \leq t \lessapprox t^\times(H)$ with $t^\times(H)\sim H^2$ diffusive smoothening of the initial
           steplike state leads to a power law behavior $\epsilon \sim t^{-\frac{1}{2}}$ (see (a)). At times 
           $t\approx t^\times(H)$ a crossover to an ultimately exponential decay of $\epsilon$ (see (b)) takes place. 
           The time scale of this exponential decay is proportional to $H^2$.}
\end{figure}

Figure~\ref{fig:B} depicts the nonstationarity parameter $\epsilon[\set{\rho}(t)]$ evaluated for the state 
$\set{\rho}(t)$ evolved from the initial state Eq.~\Ref{init} as a function of time $t$ for channel lengths
$H=100D$ (solid line), $H=50D$ (dashed line), and $H=25D$ (doted line). In Fig.~\ref{fig:B}a one identifies a 
power law behavior $\epsilon \sim t^{-\frac{1}{2}}$ within a time range $\tau \leq t \lessapprox t^\times(H)$, 
where $t^\times(H)$ denotes an (approximate) crossover time, which depends on the channel length $H$. 
The exponent is a consequence of the diffusive smoothening (see Fig.~\ref{fig:A}) of the 
initial steplike chemical potential profile: If $d(t)$ denotes the ''width'' of the interval in Fig.~\ref{fig:A}c 
where $\mu^*(z,t)$ deviates considerably from the reservoir values one expects a diffusive power law 
$d(t) \sim t^\frac{1}{2}$ as long as $d(t) < H$. Since the particle current $j(z,t)$ is determined by the gradient 
of the local chemical potential (see Eq.~\Ref{diffcurr}), which in the smoothened interval is approximately given by 
$\displaystyle\frac{\mu^*_H-\mu^*_0}{d(t)}$, one finds $\epsilon[\set{\rho}(t)] \sim d(t)^{-1} \sim t^{-\frac{1}{2}}$.
The crossover time $t^\times(H)$, which marks the end of the diffusive smoothening process, could be defined by 
$d(t^\times(H)) = H$, i.e., one expects, $t^\times(H) \sim H^2$. At times $t \gtrapprox t^\times(H)$ an ultimate 
exponential decay of $\epsilon[\set{\rho}(t)]$ is found (Fig.~\ref{fig:B}b). The decay times $\tau^*(H)$ of this 
exponential decay are $\tau^*(100D)=63260\tau$, $\tau^*(50D)=15793\tau$, and $\tau^*(25D)=3959\tau$ for
the reservoir configurations discussed in the present section. These values are proportional to $H^2$ and they are 
of the same order of magnitude as $\displaystyle\frac{H^2}{2\Gamma_{x,y,z}}$, the one-dimensional translational 
diffusion time which is expected to be related to the ultimate relaxation time scale.
 

\section{Properties of the steady state}
\label{sec:properties}

In the previous section one specific set of reservoir configurations has been fixed arbitrarily in order to 
study general features
of the temporal relaxation towards the corresponding stationary state. Here only properties of \emph{stationary 
states} and their dependence on the reservoir configurations is studied. However, the investigation is restricted
to cases with one reservoir sustaining an \emph{isotropic} structure $\set{\rho}_0$ at $z=0$ and one reservoir 
sustaining a \emph{nematic} structure $\set{\rho}_H$ at $z=H$. In contrast to the previous section both 
possible alignments of the nematic director with respect to the $z$ axis --- parallel and perpendicular --- 
are considered explicitly. We focus on the isotropic-nematic interface and on the magnitude
of the particle current as functions of the boundary conditions set by the reservoirs.

As a general qualitative feature, revealed in the previous section, the stationary state varies rapidly with 
position if it is ''close'' to the (equilibrium) bulk two-phase coexistence region (see profiles $t=10^5\tau$
in Figs.~\ref{fig:A}a and \ref{fig:A}b).
This feature is reminiscent of a free interface between an isotropic and a nematic phase at equilibrium bulk
coexistence. Here, however, the steplike profiles are nonequilibrium structures; we call them \emph{nonequilibrium 
interfaces}. There are several possibilities to describe the ''position'' $z^\m{int}$ of the nonequilibrium 
interface. Here the following three are considered: 
\begin{eqnarray}
   \rho(z^\m{int}_\rho,t=\infty)         & := & \frac{1}{2}(\rho^\m{iso}_\m{b}+\rho^\m{nem}_\m{b}), \nonumber\\
   (\max_i Q_{ii})(z^\m{int}_Q,t=\infty) & := & S^\m{nem}_\m{b},                                    \nonumber\\
   \mu^*(z^\m{int}_\mu,t=\infty)         & := & \mu^*_\m{b}.                                        \label{eq:zint}
\end{eqnarray}

\begin{figure}[!t]
   \includegraphics[width=8.5cm]{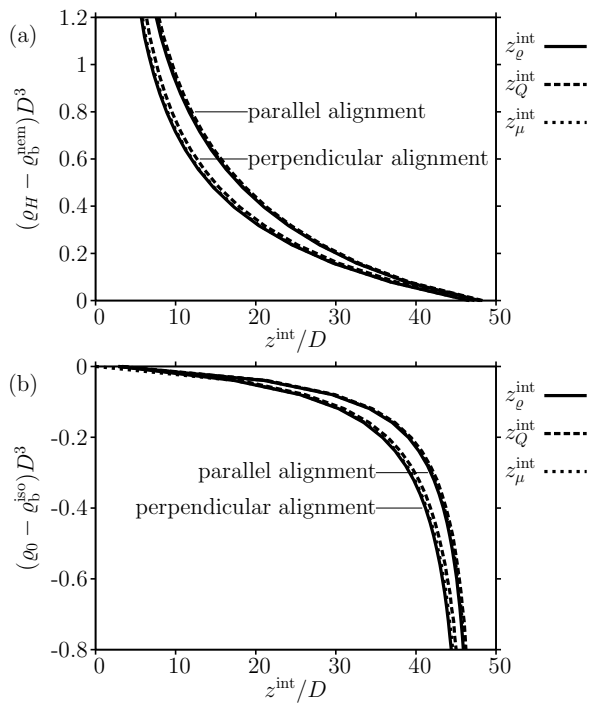}
   \caption{\label{fig:C}Position of the nonequilibrium isotropic-nematic interface $z^\m{int}$ in a channel of
           length $H=50D$ with the nematic director aligned parallel and perpendicular to the $z$ axis. In (a) the
           isotropic reservoir is fixed to $(\rho_0-\rho^\m{iso}_b)D^3 = -0.04$ whereas in (b) the nematic reservoir
           is fixed to $(\rho_H-\rho^\m{nem}_b)D^3 = 0.4$.}
\end{figure}

The different definitions of the interface position $z^\m{int}$ are displayed in Fig.~\ref{fig:C} for a channel
of length $H=50D$ and for different alignments of the nematic director of the reservoir state $\set{\rho}_H$. 
They are ordered as
$z^\m{int}_\rho < z^\m{int}_\mu < z^\m{int}_Q$ with $z^\m{int}_\mu-z^\m{int}_\rho \approx 0.1D$. Moreover, 
$z^\m{int}_Q-z^\m{int}_\rho \approx 0.3D$ for parallel and $z^\m{int}_Q-z^\m{int}_\rho \approx 0.6D$ for 
perpendicular alignment of the nematic director. These differences compared to the channel length are small,
however, hence one concludes that the three definitions for the interface position Eq.~\Ref{zint} are
equally reasonable.

Figure~\ref{fig:C}a depicts the interface position $z^\m{int}$ as a function of the density in the nematic reservoir
$\rho_H$ with the density in the isotropic reservoir fixed to $\rho_0D^3 = \rho^\m{iso}_\m{b}D^3 - 0.04$. Upon 
increasing the nematic reservoir density $\rho_H$ the nonequilibrium interface is shifted towards the isotropic 
reservoir ($z=0$). Conversely, Fig.~\ref{fig:C}b displays the interface position $z^\m{int}$ as a function of the 
isotropic reservoir density $\rho_0$ where the nematic reservoir density is fixed to 
$\rho_HD^3 = \rho^\m{nem}_\m{b}D^3 + 0.4$.
Under these conditions, the nonequilibrium interface shifts towards the nematic reservoir ($z=H$) upon decreasing
the density of the isotropic reservoir $\rho_0$. Moreover, for given reservoir densities, the nonequilibrium 
interface for parallel alignment of the nematic director is located closer to the nematic reservoir ($z=H$) 
than for perpendicular alignment.

\begin{figure}[!t]
   \includegraphics[width=8.5cm]{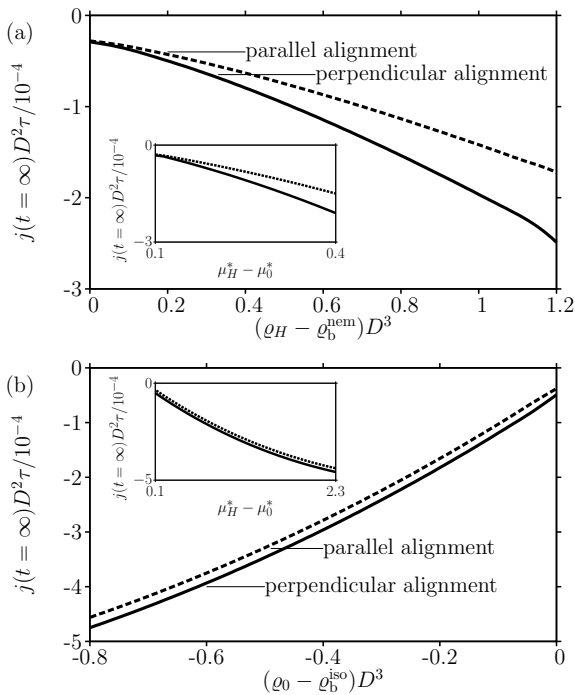}
   \caption{\label{fig:D}Stationary state current $j(t=\infty)$ in a channel of length $H=50D$ with the nematic 
           director aligned parallel and perpendicular to the $z$ axis. In (a) the isotropic reservoir density is 
           fixed to $(\rho_0-\rho^\m{iso}_b)D^3 = -0.04$ whereas in (b) the nematic reservoir density is fixed to 
           $(\rho_H-\rho^\m{nem}_b)D^3 = 0.4$. The main plots exhibit $j(t=\infty)$ as a function of (a) $\rho_H$
           and (b) $\rho_0$. An alternative representation of $j(t=\infty)$ as a function of the chemical potential
           difference of the reservoirs $\mu^*_H-\mu^*_0$ is displayed in the insets.}
\end{figure}

The dependence of the total current of the stationary state $j(t=\infty)$ on the configurations of the reservoirs
is displayed in Fig.~\ref{fig:D}. In Fig.~\ref{fig:D}a the isotropic reservoir density is fixed to 
$\rho_0D^3 = \rho^\m{iso}_\m{b}D^3 - 0.04$ whereas in Fig.~\ref{fig:D}b the nematic reservoir density is fixed to
$\rho_HD^3 = \rho^\m{nem}_\m{b}D^3 + 0.4$. The main plots exhibit $j(t=\infty)$ as a function of the reservoir 
densities $\rho_H$ (Fig.~\ref{fig:D}a) and $\rho_0$ (Fig.~\ref{fig:D}b); the insets alternatively display $j(t=\infty)$ 
as a function of the chemical potential difference of the reservoirs $\mu^*_H-\mu^*_0$.
Since $\rho_H > \rho_0$, or equivalently $\mu^*_H > \mu^*_0$, the total current is \emph{negative}. The
magnitude $|j(t=\infty)|$ increases with $\rho_H-\rho_0$ and it is larger for perpendicular than for parallel 
alignment of the nematic director. The latter observation is not surprising if one recalls $\Gamma_{x,y} > \Gamma_z$
(Eq.~\Ref{Gamma}). However, the two cases shown in Figs~\ref{fig:D}a and \ref{fig:D}b are remarkably different
in the sense, that the current difference between perpendicular and parallel alignment 
$|j_\perp(t=\infty)|-|j_\|(t=\infty)|$ is almost constant upon fixing the nematic reservoir density $\rho_H$ and 
varying the isotropic reservoir density $\rho_0$ (Fig.~\ref{fig:D}b) whereas this difference increases upon 
increasing the nematic reservoir density $\rho_H$ and fixing the isotropic reservoir density $\rho_0$ 
(Fig.~\ref{fig:D}a). Finally,
for the same reservoir density difference $\rho_H-\rho_0$, the magnitude of the stationary state current 
$|j(t=\infty)|$ in Fig.~\ref{fig:D}b where most of the channel is filled with isotropic fluid is larger than in
Fig.~\ref{fig:D}a where the channel contains predominantly nematic fluid.


\section{Discussion and Summary}
\label{sec:discussion}

In the present work nonequilibrium steady states of a fluid of platelike colloidal particles in a channel 
which connects two reservoirs sustaining bulk structures of different chemical potentials have been investigated.
A typical platelet fluid is sample A10P of Ref.~\cite{vanderKooij2000}, which consists of an aqueous
dispersion (solvent viscosity $\eta=8.9\cdot 10^{-4}\m{Pa\cdot s}$) of sterically stabilized gibbsite 
platelets of diameter $D = 165\m{nm}$. According to Eq.~\Ref{tau} the rotational relaxation time is given by 
$\tau\approx 214\m{\mu s}$. The calculated translational diffusion coefficients (see Eq.~\Ref{Gamma}) compare 
well with the measured ones of Ref.~\cite{vanderKooij2000}. The numerical results of Sec.~\ref{sec:relaxation} 
suggest that the fluid in a channel of length $H=50D\approx 8.3 \m{\mu m}$ has definitely reached the 
steady state within a time $t=10^5\tau \approx 21\m{s}$. The corresponding steady state current is
$|j(t=\infty)|D^2=7.219\cdot 10^{-4}/\tau\approx 3.4\m{s^{-1}}$. In this work the channel width is assumed to
be much larger than the particle size such that effects of the channel walls can be neglected. For a channel
width of $10\m{\mu m}\approx 61D$, say, a steady state channel current of approximately $12400$ platelets per 
second is found. As a reservoir of volume $1\ell=0.001\m{m^3}$ contains of the order of $10^{17}$ platelike 
particles, the reservoir density in an experimental realization of the setting discussed here stays constant
over any conceivable experimental time scale. 

It has been shown in Sec.~\ref{sec:relaxation} that the relaxation process of the platelet fluid towards the 
steady state is purely diffusive. It comprises a diffusive smoothening of the initial steplike fluid structure 
until the complete channel is affected, which takes a time proportional to $H^2$, followed by a structural 
relaxation corresponding to the slowest diffusion mode with a relaxation time proportional to $H^2$. The two
different regimes have been detected by a power law and an exponential decay, respectively, of the nonstationarity
parameter $\epsilon$ (Eq.~\Ref{epsilon}), which measures the inhomogeneity of the particle current within the 
channel. In the light of the rather complicated nonlocal fluid model (Sec.~\ref{sec:formalism}) involving 
translational as well as orientational degrees of freedom, the clear identification of the power law and exponential
decay regimes of $\epsilon$ as a function of time suggests that these regimes might also be easily found in
real platelet fluids.

In summary the present work studies the formation and the structure of nonequilibrium steady states in fluids
of platelike colloidal particles in a channel by means of dynamic density functional theory. Localized rapid 
changes of density (Fig.~\ref{fig:A}a) and order parameter tensor (Fig.~\ref{fig:A}b) profiles of nonequilibrium 
steady states are similar to free interfaces. The local chemical potential profile interpolates smoothly between 
the values sustained by reservoirs at the ends of the channel (Fig.~\ref{fig:A}c). The broadening of the particle 
current distribution (Fig.~\ref{fig:A}d) suggests the introduction of a parameter measuring the distance 
from the steady state. The purely diffusive relaxation process towards the steady state comprises
two regimes: a smoothening of the initial steplike structure followed by an ultimate relaxation of the slowest
diffusive mode (Fig.~\ref{fig:B}). The position of a nonequilibrium interface (Fig.~\ref{fig:C}) and the particle
current (Fig.~\ref{fig:D}) of steady states depend nontrivially on the structure of the reservoirs due to the coupling
between translational and orientational degrees of freedom of the fluid.


\begin{acknowledgments}
The authors thank Marjolein Dijkstra for access to additional computational resources.
This work is part of the research program of the 'Stichting voor 
Fundamenteel Onderzoek der Materie (FOM)', which is financially supported by 
the 'Nederlandse Organisatie voor Wetenschappelijk Onderzoek (NWO)'. 
\end{acknowledgments}



\begin{thebibliography}{00}
   \bibitem{Mourchid1995}
      A.\ Mourchid, A.\ Delville, J.\ Lambard, E.\ L\'{e}colier, and P.\ Levitz,
      Langmuir \textbf{11}, 1942 (1995).
   \bibitem{Brown1998}  
      A.\ B.\ D.\ Brown, S.\ M.\ Clarke, and A.\ R.\ Rennie,
      Langmuir \textbf{14}, 3129 (1998).
   \bibitem{Mourchid1998}
      A.\ Mourchid, E.\ L\'{e}colier, H.\ van Damme, and P.\ Levitz, 
      Langmuir \textbf{14}, 4718 (1998).
   \bibitem{vanderKooij1998}
      F.\ M.\ van der Kooij and H.\ N.\ W.\ Lekkerkerker,
      J.\ Phys.\ Chem.\ B \textbf{102}, 7829 (1998).
   \bibitem{Bonn1999} 
      D.\ Bonn, H.\ Kellay, H.\ Tanaka, G.\ Wegdam, and J.\ Meunier,
      Langmuir \textbf{15}, 7534 (1999).
   \bibitem{Brown1999}
      A.\ B.\ D.\ Brown, C.\ Ferrero, T.\ Narayanan, and A.\ R.\ Rennie,
      Eur.\ Phys.\ J.\ B \textbf{11}, 481 (1999).
   \bibitem{Levitz2000}
      P.\ Levitz, E.\ L\'{e}colier, A.\ Mourchid, A.\ Delville, and S.\ Lyonnard,
      Europhys.\ Lett.\ \textbf{49}, 672 (2000).
   \bibitem{Knaebel2000} 
      A.\ Knaebel, M.\ Bellour, M.-P.\ Munch, V.\ Viasnoff, F.\ Lequeux, and J.\ L.\ Harden,
      Europhys.\ Lett.\ \textbf{52}, 73 (2000).
   \bibitem{Abou2001}
      B.\ Abou, D.\ Bonn, and J.\ Meunier,
      Phys.\ Rev.\ E \textbf{64}, 021510 (2001).
   \bibitem{vanderBeek2003}
      D.\ van der Beek and H.\ N.\ W.\ Lekkerkerker,
      Europhys.\ Lett.\ \textbf{61}, 702 (2003).
   \bibitem{Liu2003}
      S.\ Liu, J.\ Zhang, N.\ Wang, W.\ Liu, C.\ Zhang, and D.\ Sun,
      Chem.\ Mater.\ \textbf{15}, 3240 (2003).
   \bibitem{vanderBeek2004} 
      D.\ van der Beek and H.\ N.\ W.\ Lekkerkerker,
      Langmuir \textbf{20}, 8582 (2004).
   \bibitem{Wang2005}
      N.\ Wang, S.\ Liu, J.\ Zhang, Z.\ Wu, J.\ Chen, and D.\ Sun, 
      Soft Matter \textbf{1}, 428 (2005).
   \bibitem{Cuesta1999}
      J.\ A.\ Cuesta and R.\ P.\ Sear,
      Eur.\ Phys.\ J.\ B \textbf{8}, 233 (1999).
   \bibitem{Rowan2002} 
      D.\ G.\ Rowan and J.-P.\ Hansen,
      Langmuir \textbf{18}, 2063 (2002).
   \bibitem{Harnau2001}
      L.\ Harnau, D.\ Costa, and J.-P.\ Harnau,
      Europhys.\ Lett.\ \textbf{53}, 729 (2001).
   \bibitem{Harnau2002a} 
      L.\ Harnau and S.\ Dietrich,
      Phys.\ Rev.\ E \textbf{65}, 021505 (2002).
   \bibitem{Harnau2002b}
      L.\ Harnau, D.\ Rowan, and J.-P.\ Hansen,
      J.\ Chem.\ Phys.\ \textbf{117}, 11359 (2002).
   \bibitem{Bier2004}
      M.\ Bier, L.\ Harnau, and S.\ Dietrich,
      Phys.\ Rev.\ E \textbf{69}, 021506 (2004).
   \bibitem{Harnau2004}
      L.\ Harnau and S.\ Dietrich,
      Phys.\ Rev.\ E \textbf{69}, 051501 (2004).
   \bibitem{Costa2005}
      D.\ Costa, J.-P.\ Hansen, and L.\ Harnau,
      Mol.\ Phys.\ \textbf{103}, 1917 (2005).
   \bibitem{Harnau2005}
      L.\ Harnau and S.\ Dietrich,
      Phys.\ Rev.\ E \textbf{71}, 011504 (2005).
   \bibitem{Bier2005}
      M.\ Bier, L.\ Harnau, and S.\ Dietrich,
      J.\ Chem.\ Phys.\ \textbf{123}, 114906 (2005).
   \bibitem{Bier2006}
      M.\ Bier, L.\ Harnau, and S.\ Dietrich,
      J.\ Chem.\ Phys.\ \textbf{125}, 184704 (2006).
   \bibitem{vanderBeek2006}
      D.\ van der Beek, H.\ Reich, P.\ van der Schoot, M.\ Dijkstra, T.\ Schilling, R.\ Vink,
      M.\ Schmidt, R.\ van Roij, and H.\ Lekkerkerker
      Phys.\ Rev.\ Lett. \textbf{97}, 087801 (2006).
   \bibitem{Reich2007}
      H.\ Reich, M.\ Dijkstra, R.\ van Roij, and M.\ Schmidt,
      J.\ Phys.\ Chem.\ B \textbf{111}, 7825 (2007).
   \bibitem{Bier2007}
      M.\ Bier and R.\ van Roij,
      Phys.\ Rev.\ E \textbf{76}, 021405 (2007).
   \bibitem{Evans1979}
      R.\ Evans,
      Adv.\ Phys.\ \textbf{28}, 143 (1979).
   \bibitem{Evans1989}
      R.\ Evans, in \textit{Les Houches, Session XLVIII, 1988 --- Liquides
      aux interfaces / Liquids at interfaces}, edited by J.\ Charvolin, 
      J.\ F.\ Joanny, and J.\ Zinn-Justin (North-Holland, Amsterdam, 1989), 
      p.\ 1.
   \bibitem{Evans1991}
      R.\ Evans, in \textit{Inhomogeneous fluids}, edited by D.\ Henderson
      (Marcel Dekker, New York, 1991), p.\ 89.
   \bibitem{Dieterich1990} 
      W.\ Dieterich, H.\ L.\ Frisch, and A.\ Majhofer,
      Z.\ Phys.\ B \textbf{78}, 317 (1990).
   \bibitem{Langer1971}
      J.\ S.\ Langer,
      Ann.\ Phys.\ \textbf{65}, 53 (1971).
   \bibitem{Kawasaki1977}
      K.\ Kawasaki,
      Prog.\ Theor.\ Phys.\ \textbf{57}, 410 (1977).
   \bibitem{Collins1985}
      J.\ B.\ Collins and H.\ Levine, 
      Phys.\ Rev.\ B \textbf{31}, 6119 (1985) 
      [Erratum: Phys.\ Rev.\ B \textbf{33}, 2020 (1986)].
   \bibitem{Harrowell1987} 
      P.\ R.\ Harrowell and D.\ W.\ Oxtoby,
      J.\ Chem.\ Phys.\ \textbf{86}, 2932 (1987).
   \bibitem{Boettinger2002}
      W.\ J.\ Boettinger, J.\ A.\ Warren, C.\ Beckermann, and A.\ Karma,
      Annu.\ Rev.\ Mater.\ Res.\ \textbf{32}, 163 (2002).
   \bibitem{Granasy2006}
      L.\ Gr\'{a}n\'{a}sy, T.\ Pusztai, and T.\ B\"{o}rzs\"{o}nyi, 
      in \textit{Handbook of Theoretical and Computational Nanotechnology, Vol.\ 9},
      edited by M.\ Rieth and W.\ Schommers (American Scientific Publishers, Stevenson Ranch, 2006),
      p.\ 525.
   \bibitem{Dean1996}
      D.\ S.\ Dean,
      J.\ Phys.\ A \textbf{29}, L613 (1996).
   \bibitem{Marconi1999}
      U.\ M.\ B.\ Marconi and P.\ Tarazona,
      J.\ Chem.\ Phys.\ \textbf{110}, 8032 (1999).
   \bibitem{Marconi2000}
      U.\ M.\ B.\ Marconi and P.\ Tarazona,
      J.\ Phys.: Condens.\ Matter \textbf{12}, A413 (2000).
   \bibitem{Archer2004}
      A.\ J.\ Archer and R.\ Evans,
      J.\ Chem.\ Phys.\ \textbf{121}, 4246 (2004).
   \bibitem{Schnakenberg1976}
      J.\ Schnakenberg,
       Rev.\ Mod.\ Phys.\ \textbf{48}, 571 (1976).
   \bibitem{Zwanzig1963}
      R.\ Zwanzig, 
      J.\ Chem.\ Phys.\ \textbf{39}, 1744 (1963).
   \bibitem{deGennes1993}
      P.\ G.\ de Gennes and J.\ Prost,
      \textit{The Physics of Liquid Crystals} 
      (Oxford University Press, Oxford, 1993).
   \bibitem{Cuesta1997.1}
      J.\ A.\ Cuesta and Y.\ Mart\'{\i}nez-Rat\'{o}n, 
      Phys.\ Rev.\ Lett.\ \textbf{78}, 3681 (1997).
   \bibitem{Cuesta1997.2}
      J.\ A.\ Cuesta and Y.\ Mart\'{\i}nez-Rat\'{o}n, 
      J.\ Chem.\ Phys.\ \textbf{107}, 6379 (1997).
   \bibitem{Brenner1974}
      H.\ Brenner,
      Int.\ J.\ Multiphase Flow \textbf{1}, 195 (1974).
   \bibitem{Qiu1990}
      X.\ Qiu, X.\ L.\ Wu, J.\ Z.\ Xue, D.\ J.\ Pine, D.\ A.\ Weitz, and P.\ M.\ Chaikin,
      Phys.\ Rev.\ Lett.\ \textbf{65}, 516 (1990).
   \bibitem{Xue1992}
      J.-Z.\ Xue, X.-L.\ Wu, D.\ J.\ Pine, and P.\ M.\ Chaikin,
      Phys.\ Rev.\ A \textbf{45}, 989 (1992).
   \bibitem{vanderKooij2000}
      F.\ M.\ van der Kooij, A.\ P.\ Philipse, and J.\ K.\ G.\ Dhont,
      Langmuir \textbf{16}, 5317 (2000).
\end{thebibliography}
\end{document}